\documentclass[letterpaper, 10 pt, conference]{ieeeconf}  

\IEEEoverridecommandlockouts                              

\usepackage{graphics} 
\usepackage{epsfig} 
\usepackage{amsmath} 
\usepackage{amssymb}  
\usepackage{subfig}

\usepackage{url}
\usepackage[ruled, vlined, linesnumbered]{algorithm2e}
\usepackage{verbatim} 
\usepackage{soul, color}
\usepackage{lmodern}
\usepackage{fancyhdr}
\usepackage{subfig}
\usepackage{caption}
\usepackage{graphicx}
\usepackage[utf8]{inputenc}
\usepackage{fourier} 
\usepackage{array}
\usepackage{makecell}

\SetNlSty{large}{}{:}

\makeatletter

\newcommand{\Rom}[1]{\expandafter\@slowromancap\romannumeral #1@}
\makeatother

\title{\LARGE \bf
RiskSEA : A Scalable Graph Embedding for Detecting On-chain Fraudulent Activities on the Ethereum Blockchain
}

\author{Ayush Agarwal*\thanks{*Contact Author: Ayush Agarwal, ayush.agarwal@coinbase.com}, Lv Lu, Arjun Maheswaran, Varsha Mahadevan and Bhaskar Krishnamachari
\\ Machine Learning Team, Coinbase \\
}

\begin{document}

\maketitle
\thispagestyle{plain}
\pagestyle{plain}

\begin{abstract}

Like any other useful technology, cryptocurrencies are sometimes used for criminal activities.  While transactions are recorded on the blockchain, there exists a need for a more rapid and scalable method to detect addresses associated with fraudulent activities. We present a \textit{RiskSEA} a scalable risk scoring system capable of effectively handling the dynamic nature of large-scale blockchain transaction graphs. The risk scoring system, which we implement for Ethereum, consists of 1) a scalable approach to generating node2vec embedding for entire set of addresses to capture the graph topology  2) transaction-based features to capture the transactional behavioral pattern of an address 3) a classifier model to generate risk score for addresses that combines the node2vec embedding and behavioral features. Efficiently generating node2vec embedding for large scale and dynamically evolving blockchain transaction graphs is challenging – we present two novel approaches for generating node2vec embeddings and effectively scaling it to the entire set of blockchain addresses: 1) node2vec embedding propagation and 2) dynamic node2vec embedding. We present a comprehensive analysis of the proposed approaches. Our experiments show that combining both behavioral and node2vec features boosts the classification performance significantly, and that the dynamic node2vec embeddings perform better than the node2vec propagated embeddings.

\end{abstract}


\section{INTRODUCTION}

\subsection{Background and Problem Statement}

The increasing association of cryptocurrencies to various industries has attracted wide attention from everyone. However, because of its pseudonymous nature and lack of identity tied to blockchain addresses, some criminals have used it to commit cyber crimes \cite{nadareishvili2022cryptocurrency}.  In traditional banking, in which all transactions happen on privately controlled, internal ledgers, customers cannot open accounts and transact without direct approval of the financial institution; however, in case of blockchain in which transactions happen on public ledgers and no financial institution is required to open an account, users can transact without the approval of regulated entities. It would be beneficial to compliance in the crypto space to improve visibility into transactions that happen peer-to-peer without third-party intermediaries. Moreover, as cryptocurrency transactions become more widespread, the number of transactions and wallets increase, making detection time-consuming and hard to scale.

Numerous prior studies have been aimed to identify fraudulent transactions on large financial transaction networks \cite{Lin2023TowardsUC} \cite{Kabla2022EthPSDAM} \cite{10.1145/3391195}, including Blockchain networks \cite{10.1145/3366423.3380103} \cite{10.1145/3442381.3449916}. The predominant approach in most of these studies involves modeling the patterns associated with fraudulent transactions using features derived from transaction timestamps and amounts. These models are subsequently employed to unveil previously unnoticed transactions exhibiting these distinctive patterns. To our knowledge, with a few exceptions (such as \cite{Li2022PhishingFD} \cite{Wu2019TEDGETW} \cite{Kanezashi2022EthereumFD}), the prior work has not significantly explored the use of graph embeddings for fraud detection and risk assessment. None of the prior works have focused on the scalability of such graph embeddings for the very large-scale blockchain transaction networks or adequately addressed the dynamic nature of blockchain transaction graphs as they evolve over time. 
\par

\subsection{Our Proposed Solution}

We present \textit{RiskSEA}, a scalable system for predicting the risk for all Ethereum blockchain addresses, in the form of a risk score. The risk score is normalized to be between 0 and 1, where 0 corresponds to no risk, and 1 corresponds to a very high risk of fraudulent activity associated with the address.  Our approach leverages the power of machine learning and graph analysis to identify patterns and relationships within the Ethereum network that may indicate fraudulent activity. Specifically, we use the well-known node2vec algorithm \cite{grover2016node2vec} to generate embeddings for each node in the Ethereum blockchain, and we use behavioral features based on volume, amount, and time to create more sophisticated and accurate models for fraud detection. Our experimental findings show that the combination of graph-based and behavioral features based on statistics can improve the accuracy and reliability of risk scores for Ethereum addresses. \par

The system we describe is scalable enough to be able to provide risk scores for every Ethereum address (as of September 2023, there are approximately 266 million nodes on Ethereum). For this purpose, we present and evaluate different scalability approaches.  To our knowledge it is the first such comprehensive system ever developed and deployed in practice. Further, our system is based on a dynamic variant of the node2vec algorithm \cite{8621910} and is therefore able to effectively handle the dynamic and evolving nature of transaction graphs which have new nodes and transactions added over time. 

\subsection{Node2Vec: Background and Motivation}

Node2vec is a graph-based machine learning algorithm that can be used to analyze the structure of a network. It was first introduced in a paper by Grover and Leskovec titled "node2vec: Scalable Feature learning for Networks" \cite{grover2016node2vec} and has been widely used in a variety of applications including fraud detection, recommendation systems and social network analysis. It is based on the idea of learning low-dimensional representations i.e embeddings for nodes in a graph. The algorithm works by first constructing a random walk on the graph, which is a sequence of nodes that is generated by randomly moving through the graph. It then uses the random walk to create training samples for a semi-supervised learning model, which learns the node embeddings. \par 

The node embeddings learned by node2vec algorithm capture the structural information of the graph, such as connections between nodes and relative proximity of nodes to one another. It helps in identifying patterns and relationships within a network that may not be apparent using traditional methods making it valuable for applications such as fraud detection, where ability to identify hidden patterns and relationships is critical. \par

When analyzing addresses that engage in transactions within the blockchain, we utilize node2vec embedding to generate graph based features that are provided as an input to a classifier model. Node2Vec helps us create numerical representations (embeddings) for each address based on relationships and connections it has with other addresses in the blockchain network. Malicious addresses often disperse their funds within a closely interconnected cluster of supporting addresses. By leveraging the node2vec embeddings, we can identify and analyze these dense clusters of addresses that exhibit similar transaction patterns. This enables us to detect and classify potential malicious addresses based on their association within these tightly knit groups. The embedding serves as a representation of the address’s network relationships, aiding us in understanding and distinguishing specific transactional behavior within blockchain. \par

However, node2vec embeddings have certain limitations: 1) they cannot be generated for addresses that were not part of the training data, 2) generating node2vec embedding for large scale transaction graphs with hundreds of millions of nodes is computationally challenging, 3) it requires re-computation to handle the evolving nature of blockchain transaction graph which is very time-consuming. To tackle these challenges, we consider two approaches. The first approach propagates node2vec embeddings from the nodes involved in the training to all other nodes. The second approach implements an incremental approach to train the node2vec model to handle the dynamic nature of blockchain and generate embeddings for the addresses. In our incremental approach each training iteration only needs to consider incremental changes in the transaction graph over a period of time, and embeddings are only updated for changed regions since the last training. The incremental training approach is inspired by the paper from Mehdavi et al. titled "dynnode2vec: Scalable Dynamic Network Embedding" \cite{8621910} with few modifications to efficiently generating random walks on large graphs. With both approaches we are able to successfully generate embeddings for a complete spectrum of addresses in blockchain, and with the second approach we are also able to handle the evolving nature of the blockchain transaction graph.  \par

\subsection{Our Contributions}

The main contributions of our work are as follows: \\

1. Risk Scoring System: We present \textit{RiskSEA}, a novel risk scoring system designed to efficiently and effectively generate normalized risk scores for Ethereum blockchain addresses, to predict their likelihood of being involved with fraudulent activity. The system employs a supervised machine learning approach involving a soft binary classification model, with outputs corresponding to the risk score. For training and evaluation, we use an extensive set of datasets for obtaining the ground-truth binary risk labels, including both addresses known to have engaged in malicious behavior as well as benign addresses.  \par

2. Comprehensive Set of Features:  The model uses a comprehensive feature set combining both behavioral and graph-based features. The behavioral features aim to capture the transactional behavior pattern of an address based on transaction amount and timestamps. The graph-based node2vec feature we use captures topology of the Ethereum transaction graph (where addresses are nodes and directed edges correspond to transactions between them).  \par

3.  Demonstrating Benefit of Graph Features: We show through an ablation study that graph-based features  a) provide better classification results compared to behavioral features, and b) when combined with behavioral features provide a better result than either behavioral or graph-based features in isolation. \par

4. Scalable approaches: Motivated by the goal to generate risk scores for all Ethereum addresses (hundreds of millions of them), we develop, implement and evaluate techniques to generate node2vec embedding for the entire Ethereum transaction graph. We compare two main approaches: 1) an embedding propagation approach that computes node2vec embeddings for a subset of addresses and propagates them to other addresses in a specified manner, and 2) an incremental dynamic node2vec approach, that is capable also of handling the dynamic and time-evolving nature of blockchain transaction graphs. In conjunction with the latter, we also present a novel horizontal scalable random walks approach that allows us to efficiently implement incremental training for the dynamic node2vec model. To our knowledge this is the first-ever system for risk scoring that has been implemented and deployed at the scale of the entire Ethereum transaction graph and capable of being incrementally updated over time as the transaction graph evolves. 
\par

5. Evaluations of proposed scheme: We show that the dynamic node2vec approach is both computationally scalable and capable of handling dynamic transaction graphs in an incremental manner, and results in better risk score predictions. We provide a comprehensive analysis evaluating the efficacy of the dynamic node2vec embedding in terms of the risk prediction under various hyperparameters of the node2vec model.

\subsection{Organization of the Paper}
The remainder of the paper is as follows. In Section 2, we present related work on the problem of detecting fraudulent transactions in blockchain and methods using graph based embeddings and GNN models. In Section 3, we introduce our overall approach for generating risk scores of addresses based on a combination of statistical features of the addresses and address embedding generated using the node2vec model. In Section 4, we present the methodology for propagating the embedding to entire set of addresses and it’s drawbacks, and about horizontally scalable random walks for efficiently training a node2vec model, an incremental training approach to generate address embeddings, as well as a performance evaluation of the address embedding with varied node2vec parameters and how it is able to tackle the evolving nature of the transaction graph. Finally, Section 6 concludes this study and discusses the scope of future work and different applications.

\section{RELATED WORK}

Fraud detection in the Ethereum transaction network is one of the hot topics due to its social importance and the availability of Ethereum transaction data publically. Financial transaction networks  usually have edge attributes - timestamp and amount. These attributes are assumed to be the key to fraud detection. Many fraud detection methods have been proposed using transaction edge attributes based on this assumption. However, some proposed models and algorithms are used with arbitrary features, which may not apply to other transaction network datasets or other types of fraud detection tasks. \par

With the development of graph embedding and graph neural network research, phishing detection methods \cite{wu2020phishers} using graph embedding have been proposed. The advantage of using graph embedding and GNN models is to capture features of suspicious account node. \par

Kilic et al’s research on fraud detection using ML \cite{Kili2022FraudDI} proposes an automated system for detecting blockchain addresses through the construction of transaction graphs and extraction of address features including global features like pagerank, using standard machine learning algorithms. Wu et al. proposed a transaction embedding model named trans2vec \cite{9184813}, incorporating the amount and timestamp properties of the transaction edges. They compared trans2vec with state-of-the art embedding algorithms, and it achieved better model performance with time and amount of features. While creating account based embedding vectors, it applies random walks with the amount and time-biased sampling. \par

Kanezakshi et al’s research on ethereum fraud detection with heterogeneous graph neural networks \cite{Kanezashi2022EthereumFD} compared the model performance of GNN models on the actual Ethereum transaction network dataset and phishing reported label data and showed that heterogeneous models had better model performance than homogenous models.\par

Chen et al. proposed another phishing detection model \cite{Li2022PhishingFD} based on GCN and autoencoder. It samples subgraphs by random walk and applies node embeddings and a GCN model to incorporate spatial structures and node features. The proposed model performed better than other traditional embedding methods. However, some node features from transaction data are determined arbitrarily, and only the GCN model is used to extract the structural features. Tan et. al also proposed \cite{Tan2023EthereumFB} a framework for detecting fraudulent transactions, which involves the use of a transaction behavior-based network embedding algorithm and GCN to classify ethereum addresses as legitimate or fraudulent. \par

Lin et al. modeled the Ethereum transaction network as a weighted temporal graph and a Temporal Weighted Multidigraph Embedding \textit{T-EDGE}~\cite{Wu2019TEDGETW} to incorporate temporal and weighted transaction edges. In T-EDGE, it extends the edge probabilities to be visited in the random walk by transaction amount and interval. Xie et al. also modeled it as a temporal-amount snapshot multigraph and extended the random walk named temporal-amount walk \textit{TAW}~\cite{Xie2021TemporalAmountSM} to generate embeddings. Construct a transaction subgraph network \textit{TSGN}~\cite{Wang2021TSGNTS} with a learnable transaction weight mapping function and directed-TSGN to be aware of the edges. The model performance of TSGN is better than the baseline method in other Ethereum datasets.\par

Overall, the literature on fraud detection on detecting fraudulent addresses on ethereum blockchain shows that there is a variety of embedding approaches for nodes in a transaction graph that can be used to identify and prevent fraudulent activity in the system. However, there is little or no research in terms of a scalable approach which can be applied to each address in the ethereum blockchain in an efficient way. Our work addresses this gap in the literature.\par

\begin{figure*}[ht]
    \centering
    \includegraphics[height=2.55in, width=7.2in]{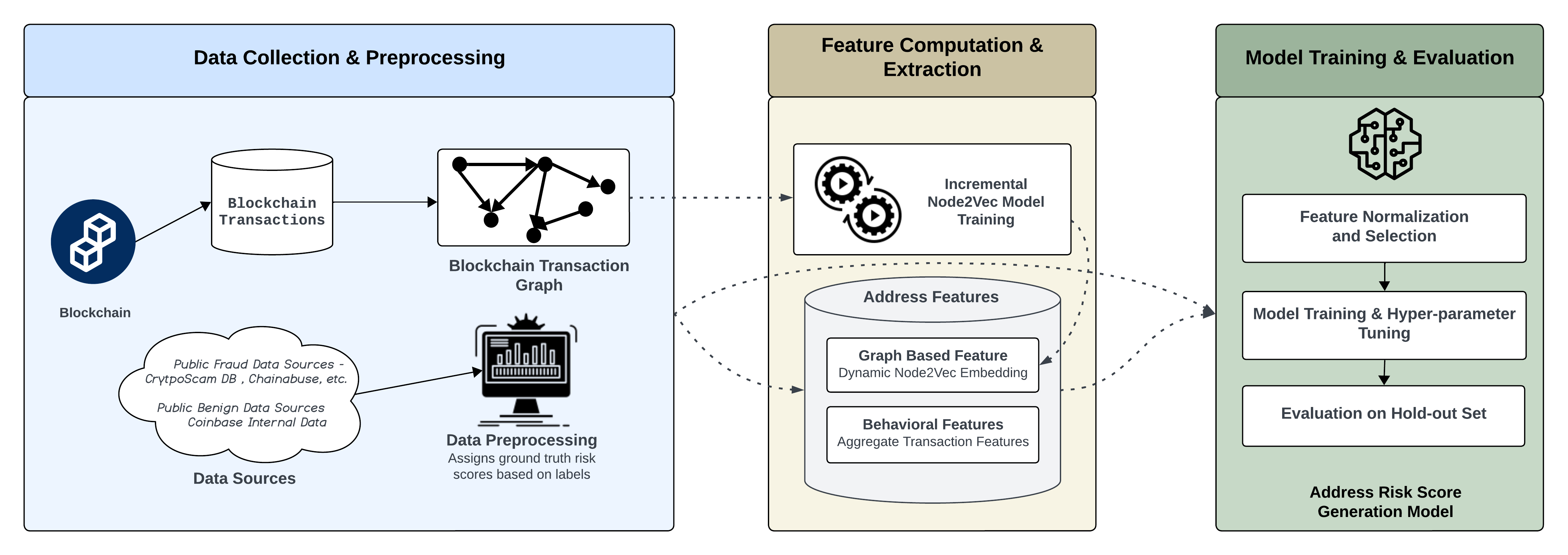}
    \caption{Risk Scoring of Ethereum Blockchain Addresses \textbf{\textit{(RiskSEA)}}}
    \label{fig:2}
\end{figure*}

\section{Risk Scoring of Ethereum Addresses \textit{\textbf{(RiskSEA)}}}

This section briefly describes the overall pipeline in our \textit{RiskSEA} system for generating risk scores of ethereum blockchain addresses. This pipeline includes the data collection and processing, feature computation and extraction, and model training and evaluation.

\subsection{Data Collection and Preprocessing}

When working on problems where security is a primary concern, it is crucial to carefully analyze the data we ingest  and train the models on. It is particularly crucial to ensure a balanced representation of "bad" actors, i.e., addresses involved in malicious activities, due to the significant imbalance between such actors and "good" actors.  To achieve this, a curated dataset was created by evaluating various sources for their genuineness and quality of labels and their feasibility to be incorporated into our systems. \par
\vspace{2mm}

In the data preparation step, we considered different sources of data \footnote{These data sources are not revealed because of security concerns}. The data points collected from various sources provide labels associated with an address highlighting the form of activity the address is involved in, examples of such labels include - phishing, scam, theft, miner, exchange, etc.  Based on these labels, risk class is assigned to these ground truth addresses. 

We have labeled addresses into two risk classes: 
\begin{itemize}

\item  Risk-1/“Bad” actors: addresses involved in illegal activities
\item Risk-0/“Good” actors: addresses belonging to exchanges, miners, government, etc.

\end{itemize}

For addresses having different risk scores coming from different sources, a high risk score is given precedence. 

\begin{table}
    \centering
    \begin{tabular}{c|c}
         \textbf{Data Sources} & \textbf{Number of Data Points}\\ 
         \hline
         \hfill
         Coinbase Internal Data\footnote{Coinbase random sample of KYC'ed customer wallet addresses and blocklisted addresses}& 43334\\ 
         External Data Sources& 85678\\ 
    \end{tabular}
    \label{table:1}
    \caption{Distribution of Internal and External Data Sources}
\end{table}

\subsection{Feature Extraction}
Our analysis of the Ethereum transaction graph revealed distinct spending patterns among addresses of different risk categories. To leverage this insight, we provide our classifier model with a comprehensive summary of an address's activity. By considering transactional behavior and patterns, we can better understand the characteristics associated with different risk levels, enabling more informed assessments and analysis.  However these transactional features can not fully capture the transaction pattern or structural pattern of different blockchain addresses, so to capture the  structural information, we use a graph based feature Node2Vec. It captures topology of an address with respect to other addresses, helps in identifying patterns and relationships within a network that may not be apparent using traditional methods.

\subsubsection{Behavioral Features}
These features capture the behavioral transaction patterns among different addresses, these are features based on timestamp, amount, and frequency of the incoming/outgoing transactions for a particular address. \par 

We created two buckets of behavioral features - one based on Ethereum transactions  of a particular address and the other based on ERC-20 token based transactions performed by an address. This helps in acquiring valuable insights from both types of transactions, capturing both allows us to utilize the trading patterns linked to different Ethereum-based tokens as a potential indicator to assess address risk score.

\subsubsection{Node2Vec Embeddings}

Node2Vec is a graph-based machine learning algorithm that can be used to analyze the structure of a network \cite{grover2016node2vec}. It is based on the idea of learning low-dimensional representations i.e embeddings for nodes in a graph. The node embeddings learned by Node2Vec algorithm capture the structural information of the graph, such as connections between nodes and relative proximity of nodes to one another.  \par

When analyzing addresses that engage in transactions within the blockchain, we utilize Node2Vec embedding to generate graph based features for our classifier model. Node2Vec helps us create numerical representations (embeddings) for each address based on relationships and connections it has with other addresses in the blockchain network. Embeddings generated by Node2Vec are particularly valuable because they capture the topology of the graph. Malicious addresses often disperse their funds within a closely interconnected cluster of supporting addresses. By leveraging the Node2Vec embeddings, we can identify and analyze these dense clusters of addresses that exhibit similar transaction patterns. This enables us to detect and classify potential malicious addresses based on their association within these tightly knit groups. The embedding serves as a representation of the address’s network relationships, aiding us in understanding and distinguishing specific transactional behavior within blockchain.

However, generating node2vec embedding for entire set of blockchain addresses is challenging and have certain limitations 

\begin{enumerate}
    \item they cannot be generated for addresses that were not part of the training data, which is required since blockchain is an evolving graph of addresses and transactions.
    \item Generating node2vec embedding for large scale transaction graphs is challenging. To put this in context, as of 20th Sep, 2023 Ethereum transaction graph had approximately 266M nodes. 
    \item Node2vec require re-computation of embeddings to  handle the evolving nature of the blockchain transaction graph which is very time-consuming.
    
\end{enumerate}

\vspace{2mm}

To tackle these scalability challenges, we explored multiple approaches to scale the node2vec embeddings to the entire set of blockchain addresses. In the section IV we present approaches for scaling node2vec embedding and the overall performance of the risk scoring model in capturing fraudulent addresses.

\subsection{Risk Scoring Model}

In our risk scoring model, we employed three major sets of features which can capture the transactional pattern of an address.  We created two sets of behavioral features based on features defined in Table 1, first only based on Ethereum transactions of an address and second based on ERC-20 transactions for a particular address.  Apart from these behavioral features we created a graph based feature - node2vec embedding, which captures topology of the graph.  In order to classify an address into a specific risk class, we utilize a random forest classifier model. We combined all three sets of features and the resulting vector is used as input to the classifier model. This approach enables us to leverage the combined power of Node2Vec embeddings and transactional data for accurate risk classification of addresses. 

To assess the performance of our risk scoring model and conduct a comparative analysis of various models, we have adopted a set of baseline metrics, including Precision, Recall, and the area under the precision-recall curve (PR-AUC). These metrics enable us to evaluate how effectively the model identifies and labels known malicious addresses. We evaluated the model's performance on a diverse test dataset, which comprises addresses from a variety of sources. The diverse test dataset allows us to gauge the model's accuracy in identifying malicious addresses from a wide range of origins.

Overall, the combination of different behavioral features and node2vec embeddings has performed very well in capturing fraudulent addresses in the blockchain. In section IV we present the performance of the risk score generation model with different scaling node2vec embedding approaches. We also presented that the node2vec feature has a very high feature importance compared to other features using the risk scoring model defined in this section for capturing fraudulent addresses.


\par

\begin{figure*}[th]
    \centering
    \includegraphics[height=1.8in, width=6.3in]{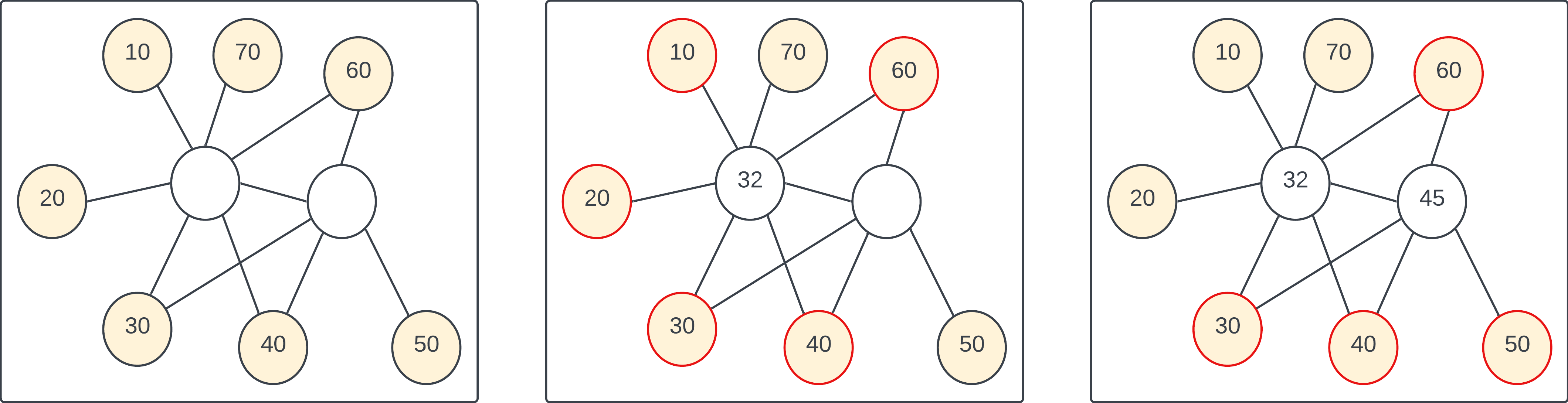}
    \caption{Embedding Propagation Methodology. Here the colored nodes represent addresses having node2vec embedding (a 1-D vector here) with value equal to the number assigned to them. For the nodes under consideration its randomly picked neighbors are highlighted in bold red outlines.
}
    \label{fig:3}
\end{figure*}

\section{Scaling Node2Vec Embeddings of Blockchain Addresses}

Blockchain ecosystem characterized by its decentralized and immutable ledger has grown exponentially over the years with cryptocurrencies like Bitcoin and Ethereum at the forefront. As of 20th Sep 2023, Ethereum boasts approximately 266M addresses and Bitcoin has approximately 1.2B addresses. These numbers reflect the scale and complexity of blockchain networks which presents a unique challenge for generating node embeddings for address. Despite the increasing significance of blockchain, there has been a noticeable gap in research when it comes to scalable approaches for generating embeddings in blockchain graphs. We have used node2vec to generate blockchain node embeddings, but there are challenges in generating node2vec embedding for large scale evolving blockchain graphs. Node2Vec is a transductive node embedding algorithm i.e it needs the list of all nodes to be available at the time of training to learn their embedding. This transductive nature restricts its direct applicability to generate embeddings for new blockchain addresses that were not part of the initial training set. We address these limitations by proposing approaches that adapt to the scalability and the evolving nature of the blockchain transaction graph and enable generation of embeddings for every node within the network.

\subsection{Embedding Propagation Methodology}

In embedding propagation methodology we tackle the scalability limitations by propagating the embeddings from a known set of addresses for which the node2vec embeddings are computed to a unknown set of addresses with missing embeddings. We start by first training a node2vec model on a set of ~ 30M addresses\footnote{ We performed the training on NVIDIA V-100 GPU and 30M addresses embedding entirely using the GPU Memory. This address set count can vary on different machines. 
} and generate node2vec embeddings, this forms our core node2vec address embeddings. To come up with core ~30M addresses, we used our training dataset where we included all the addresses from ethereum blockchain which have transacted with any addresses in the training set and randomly sampled a set of ~30M addresses from a broader set of addresses.

Once we have the core node2vec address embeddings, we propagate the embeddings. To propagate the embeddings, for a particular address X we identify all the addresses in the core node2vec address set that address X has directly transacted i.e one-hop addresses. From these addresses, we randomly sample five addresses and aggregate their node2vec embeddings to create an embedding for address X. This approach allows to efficiently extend embeddings to unrepresented addresses that were not part of the initial node2vec model training.  The propagation can be iteratively extended to multiple-hops, but for the sake of preserving the quality of the node2vec embeddings we choose to limit the embedding propagation to one-hop.

Propagated embeddings could not be generated for the entire set of addresses since it requires an edge from the core set of node2vec addresses. The propagation algorithm generated embeddings for 78\% addresses.  We also found that on average of all the addresses that were transacting daily, we were able to generate embeddings for 80-85\% of the addresses. With embedding propagation methodology, even though we can produce embeddings for the majority of addresses in the network, over time as the ethereum network evolves without repetitive address sub-sampling, training and propagation process we will run into multiple challenges.

\begin{table}
\centering
\renewcommand{\arraystretch}{1.4}

\begin{tabular}{|p{3.5cm}|c|c|c|c|}
\hline
\textbf{Model Type / Metrics}& \textbf{Precision}& \textbf{Recall}& \textbf{F1-Score}\\
\hline
Behavioral Features& 0.803 & 0.65 & 0.718 \\
\hline
Propagated Node2Vec Embedding& 0.798 & 0.686 & 0.738 \\
\hline
Propagated Node2vec Embedding + Behavioral Features& 0.908 & 0.801 & 0.851 \\
\hline

\end{tabular}
\caption{Result of \textit{RiskSEA} with different combination of behavioral features and propagated node2vec embeddings}
\label{tab:1}
\renewcommand{\arraystretch}{1}
\end{table}

\begin{algorithm}
\DontPrintSemicolon
\SetKwProg{Fn}{CoreNode2VecEmbedding}{}{end}
\Fn{train\_embedding\_model()} {
  \textbf{\textit{\#Generate embedding for CoreNode2VecSet}} \\
  \textbf{\textit{CoreNode2VecSet}} {
    \textit{Randomly sampled set of 30M address from address with transactions from training set}
  }
}
\BlankLine
\SetKwProg{Fn}{Function}{ is}{end}
\Fn{propagate\_embedding(address)} {
  \textbf{\textit{\# Generates Propagated embedding for address }} \\
  \ForEach{X in CompleteAddressSet} {
    \textit{AddrList} $=$ {
        Find address from \textit{CoreNode2VecSet} that has directly transacted with \textit{X}
    } \\
    \textit{NeighAddr} $=$ RandomSample(\textit{AddrList, 5}) \\
    \textit{PropogatedEmbedding} $=$ Aggregated embeddings of addresses in \textit{NeighAddr} \\
  }
}
\label{algo:1}
\caption{Embedding Propogation Algorithm}
\end{algorithm}

\subsection{Drawbacks of Embedding Propagation Methodology}

Specifically we will run into the following challenges:

\begin{enumerate}

    \item Propagation will make connected addresses have very similar or identical embeddings, which makes it impossible to compute embeddings to capture structural similarities. However, structural similarities are quite important for fraud detection problems as malicious users tend to have similar interaction/link structures with different users.

    \item Embeddings for addresses in newly developed sub regions will become stale/unavailable as the majority of edges in those new sub regions were not considered during embedding training. For example, when new smart contracts are deployed and start to gain more popularity, their embeddings can only be generated and updated through interacted addresses that were included in embedding training.

    \item When existing addresses that present in training start to develop new linkage patterns, their embeddings will not be updated since they are only computed during training.

\end{enumerate}

\vspace{2mm}

Even though we can perform regular re-training of the node2vec model to tackle the above challenges, we will still face few issues: 

\begin{enumerate}
    \item Due to computational resource limitations, the training dataset for Node2Vec is constrained in terms of the number of addresses and links between them. This restriction arises from the necessity of loading the entire network into memory, particularly during the execution of random walks. Consequently, as the network expands, an increasing number of addresses become inaccessible for embedding computation. Although it is possible to enhance coverage by iterating the propagation process, the quality of embeddings for nodes situated farther from those included in the initial training set is expected to diminish.

    \item Retraining on a large graph is time \& resource consuming, which will limit how frequent the retrain will be conducted.

    \item Retrain will cause node embeddings to change, which leads to much higher cost if we want to store embeddings in hot data storage for fast serving as we need to repopulate embeddings for all addresses. At the same time, any downstream models will also need to be retrained as embedding space has changed after retrain.

\end{enumerate}

To overcome these challenges, we have devised an alternative solution known as \textit{Dynamic Node2Vec Embedding Generation}. This approach is designed to address the dynamic nature of the blockchain network by efficiently generating node2vec embeddings for the entire address set. Employing an incremental methodology, the model is trained in a manner that handles the evolving nature of the blockchain transaction graph, incorporating new addresses or transactions seamlessly into the embedding generation process.

\subsection{Dynamic Node2Vec Embedding Generation on Blockchain \textbf{(DynN2VChain)}}

To overcome the challenges faced by \textit{
Embedding Propagation Methodology} and to handle the evolving nature of the blockchain transaction graph, we need to have a graph learning algorithm that supports incremental training such that each training iteration only needs to take incremental graph information as input and embeddings are only updated for addresses in changed regions since last training. We found that \textit{Dynamic Node2Vec} learning algorithm [2] effectively provides such characteristics. Dynamic node2vec is a scalable dynamic network embedding for large evolving networks. It modifies the well known static embedding method, node2vec by employing the previously learned embedding vectors as initial weights for skip-gram model. In addition, it utilizes evolving random walks for updating the trained skip-gram model where walks are generated only for nodes that have changed in consecutive timestamps. However, traditional libraries for computing random walks load the entire graph into memory which limits the size of the graph due to memory constraint. To overcome this, we developed a novel MapReduce approach that can compute random walks in a horizontally scalable manner.

\begin{figure*}[t]
    \centering
    \includegraphics[height=4.1in, width=6.2in]{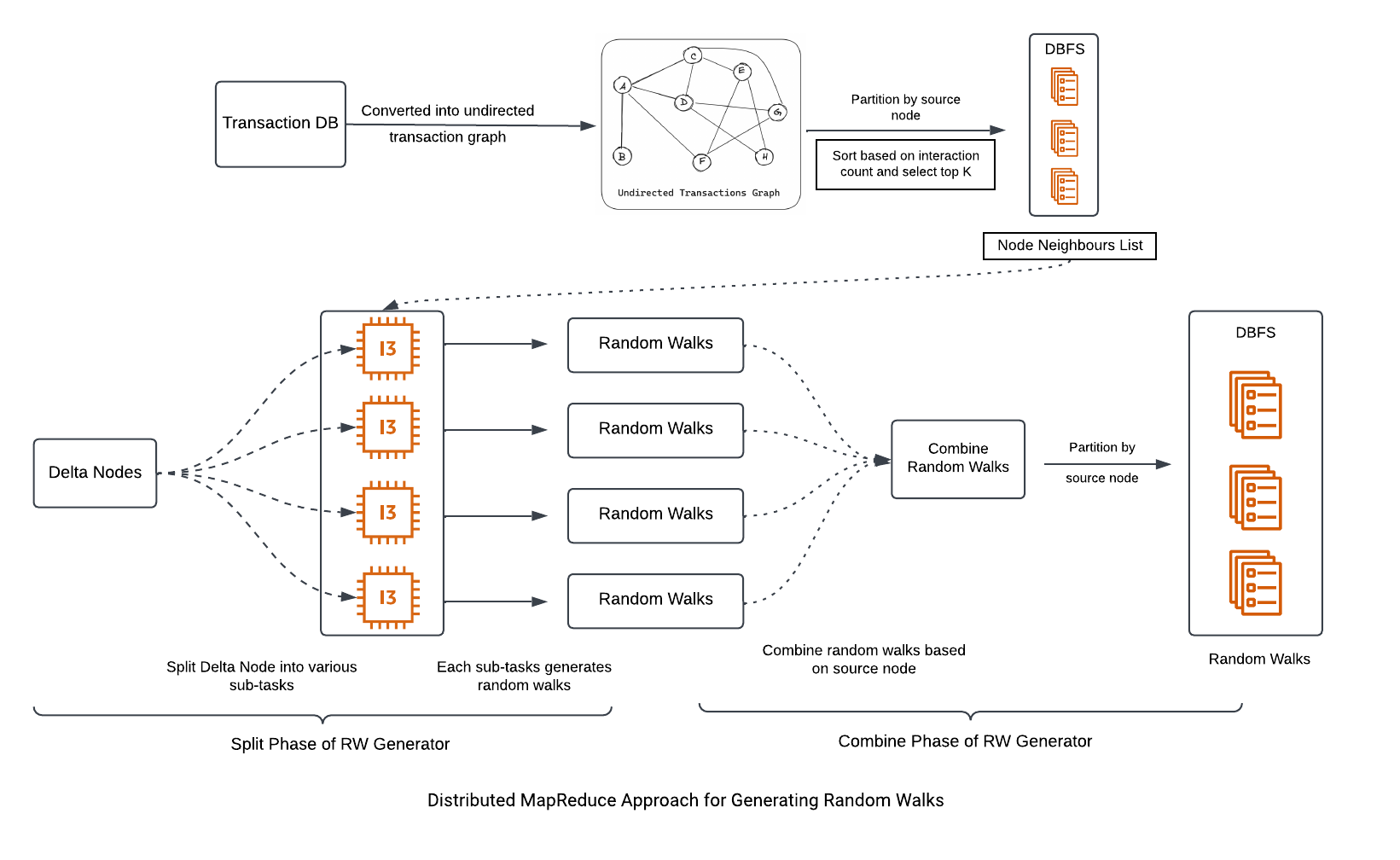}
    \caption{Distributed MapReduce Approach for Generating Random Walks}
    \label{fig:3}
\end{figure*}

\subsection{Dynamic Node2Vec}
Dynamic node2vec has two main subcomponents - node2vec random walk and skip gram model. Node2Vec random walks is a flexible random walk which samples neighborhoods of a source node by Breadth-first sampling (BFS) and Depth-first sampling (DFS). Node2Vec generates a corpus by sampling a number of fixed lengths starting at each vertex. Then, the Skip-gram uses these random walks to learn the representation vector of each node. For a dynamic network G = G1, G2, … GT, it runs the static node2vec for the first graph G1 separately, extracts the embedding vectors and keeps the structure of trained Skip-gram for next timestamp. For all other subsequent timestamps 2, … T, the following two steps i.e Evolving Walk Generation and Dynamic Skip-gram Model are performed between two consecutive timestamps. 

\subsubsection{Evolving Walk Generation}
In the static node2vec, random walks are generated independently for each timestamp for all nodes which is a very time-consuming process. In dynamic node2vec, random walks are generated only for evolving nodes instead of generating random walks for all nodes in the current timestamp. Therefore, new random walks from changed regions in the graph can efficiently update the embedding vectors according to temporal evolution of networks over time. However, in the worst case evolving nodes include all nodes as all nodes might have changed in the timestamp which in case of blockchain is very rarely possible or not possible. 

\subsubsection{Dynamic Skip-gram Model}
In the natural language processing domain, several dynamic word embeddings have been proposed to track word evolution such as new words are created, and some words die out throughout time \cite{Bamler2017DynamicWE}\cite{May2017StreamingWE}. For learning the representation vectors at timestampt they train the Skip-gram by initializing word vectors obtained from previous timestampt-1.  In this dynamic Skip-gram model, the vocabulary set is updated and the Skip-gram is retrained by new documents in new timestamps. In network embedding domains, dyn-node2vec also takes advantage of dynamic Skip-gram model for obtaining embedding vectors at time t and uses the pre-trained Skip-gram model as initial weight for further training. In order to do that, first the vocabulary set of Skip-gram is updated according to new evolving walks. Then, Skip-gram is trained by new evolving walks generated on the evolving nodes.

Using Dynamic Node2Vec, we developed an incremental approach to generate the embeddings for evolving blockchain transaction graphs and we consider addresses that have transacted since the last run of training as delta nodes and evolving walks for those nodes as delta walks. To compute delta walks, even though we only need to conduct walks for delta nodes, however, each random walk will still need to be conducted against the whole graph. For example, A is used to be only connected to B. In the new iteration, it develops a new connection to C. When computing which node from A to walk to, both B and C need to be considered in the randomized decision. Therefore the entire graph needs to be accessible when making random walk decisions. 

Traditionally, libraries that support random walk will first load the entire graph into memory, which limits the size of the graph due to memory constraint. Then to speed up random walk computation, parallelization is usually adopted to conduct random walks on different nodes concurrently. However the degree of parallelism is still constrained by the amount of computation resources that a single machine has. To overcome this, we developed a horizontally scalable approach for generating random walks.

\subsection{Horizontally Scalable Random Walks on Blockchain}

To overcome the above limitations, we developed a novel MapReduce [3] approach that can compute random walks in a horizontally scalable manner. For each timestamp, we first store the graph as node pairs on a distributed file system partitioned by source node then to avoid data skew issues\footnote{Hub nodes tend to have extremely high degree of connections} for each source we considered top K neighbors\footnote{For Ethereum Transaction graph, we consider K = 200} sorted based on interaction count. The node pairs are stored to get a list of neighboring nodes for source nodes, we call it NodeNeighborList. At each random walk step, we evaluate and pick destination nodes based on the node2vec hyper params for different delta nodes in parallel across multiple machines as a split-apply-combine procedure. In the random walk step, first delta nodes are splitted into several subtasks, each sub task iterates through assigned delta nodes and for each delta nodes, it loads a small file that contains all the neighboring nodes from the distributed file system that stores graph data, then conducts a randomized selection process according to the node2vec hyperparameters. Random walks are then combined based on their source node and saved to a distributed file system.

\begin{algorithm}
\DontPrintSemicolon
\SetKwProg{Fn}{Function}{ is}{end}
\Fn{node\_neighbor\_list()} {
  \textbf{\textit{\#Stores the graph as node pairs on distributed file sytem partitioned by source node}} \\
  
  \ForAll{source\_nodes}{
    \textit{neighbor\_nodes} = Consider top K neighbors sorted based on interaction count \;
  }
}
\BlankLine
\SetKwProg{Fn}{Function}{ is}{end}
\Fn{generate\_random\_walks(transaction\_graph)} {
  Delta Nodes are splitted across various subtasks \;
  \ForEach{task:  sub\_tasks} {
        \ForEach{node: source\_nodes}{
                neighbor\_nodes = \textit{node\_neighbor\_list()}
                \textbf{\textit{\#Performs randomized selection on neighbor\_nodes  based on hyper-parameters}\\
                }
                $r\_node$ = \textit{random\_node\_selection(neighbor\_node)\\}
        }
\textbf{        \#Random nodes combined from different sub\_tasks into partitions
}       \textit{ save\_to\_distributed\_file\_system()}
  }  
  }
\label{algo:2}
\caption{Distributed Map-Reduce Approach for Generating Random Walks}
\end{algorithm}

Using the distributed approach for random walks, we could generate random walks for large graphs in a fast and time-efficient manner on blockchain transaction graphs. After initial training, we perform an incremental training daily where it only needs to generate random walks on the delta nodes and feed those random walks to model for retrain. For nodes that are involved in walks of delta nodes i.e delta walks, their embedding in training will be initialized with values from previous iteration and updated through retraining. For nodes that are not involved in the delta walks, their embedding will not be updated in retraining.

\begin{figure*}[ht]
    \centering
    \includegraphics[height=3.2in, width=7.0in]{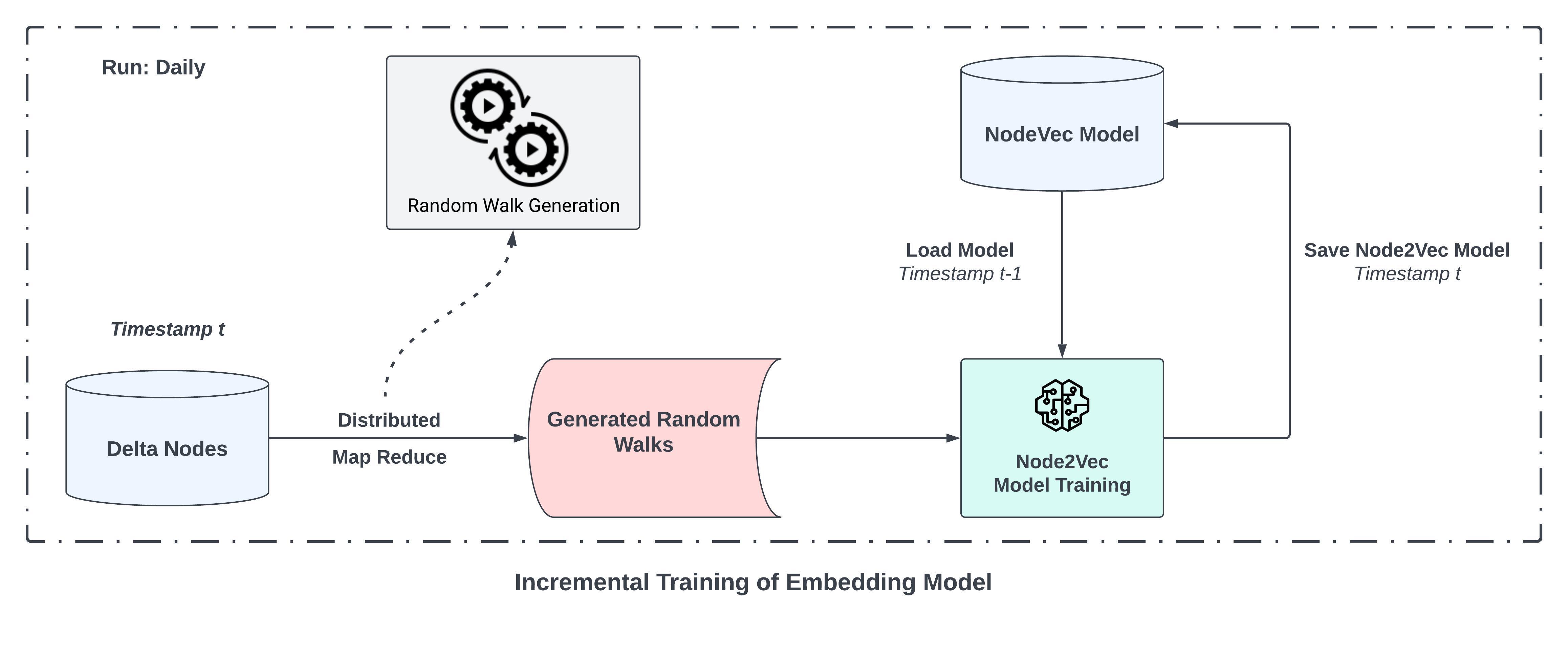}
    \caption{Incremental Training of Node2Vec Embedding Model}
    \label{fig:4}
\end{figure*}

\subsubsection{Hardware and Software Setup}

In our experiments, we have completed all experiments on Databricks clusters, we used CPUs and PyTorch as suitable hardware and libraries for deep learning to train exhaustively various GNN models and compare and evaluate their performance. 

To implement the distributed random walks we have used a databricks cluster with driver node as i3.8xlarge and similar configuration for other nodes in the cluster. To generate embeddings, we have used memory optimized clusters r5.16xlarge\footnote{R5 instances are next-gen memory optimized instances of AWS EC2} as the driver node and r5.xlarge as the other node.

We used Pytorch (1.7.1) and gensim (4.2.0) libraries to train the Node2Vec embedding models on the CPU.

\begin{algorithm}
\DontPrintSemicolon
\SetKwProg{Fn}{Function}{ is}{end}
\Fn{incremental\_dynN2Vchain\_training()} {

\textbf{\#Find nodes involved in transaction since last run $t-1$} \\
delta\_nodes = \textit{ find\_delta\_nodes}()

random\_walks = \textit{generate\_random\_walks}()

\textbf{\# Load node2vec model at timestamp $t-1$ and perform model training} \\
n2v\_model = \textit{load\_node2vec\_model}() \\

n2v\_model = \textit{node2vec\_model\_training}() \\
\textit{save\_n2v\_model}(timestamp\_t)  \\
}
\label{algo:3}
\caption{Incremental Training of Node2Vec Embedding Model}
\end{algorithm}

\subsection{Hyper-parameter Tuning of DynN2VChain Model}

For hyper-parameter tuning of the dynamic node2vec model , instead of generating the embedding for complete Ethereum transaction graph, we created a subgraph of Ethereum transaction graph by randomly sub-sampling 30M addresses which have transacted from the known ground truth addresses in our dataset. We performed the evaluation of node embeddings by generating the risk score using the \textbf{\textit{RiskSEA}} approach as described in selection III. \par

\vspace{1mm}

We performed the \textit{DynN2VChain} model training by varying all the different parameters of node2vec i.e number of walks, length of walks, Return parameter ($p$) and Inout parameter ($q$) :

\textbf{Number of Walks ($r$) - }The number of random walks to perform from each node. It influences the exploration of the graph and helps in capturing different facets of the neighborhood around a node.

\textbf{Length of Walks ($l$) - }The number of nodes to visit in each random walk. It controls the length of the node sequences generated during the random walks

\textbf{Return Parameter ($p$) }-  The likelihood of immediately revisiting the previous node in the random walk. Setting it to higher value ensures that we are less likely to sample an already visited node. On the other hand if p is low, it would lead the walk to backtrack a step and this would keep the walk close to the starting node .

\textbf{Inout Parameter}\textbf{ ($q$):} The likelihood of moving away from the previous node in the random walk. It allows the nodes to differentiate between inward and outward nodes, if q > 1 the random walk is biased towards nodes close to the starting node. Such walks obtain a local view of the underlying graph with respect to the start node. In contrast if q < 1, the walk is more inclined to visit nodes which are farther away from the start node, such behavior is reflection of DFS which encourages outward exploration.

Intuitively, parameters p and q control how fast the walks explore and leave the neighborhood of starting node u. It allows the walk generation process to interpolate between BFS \& DFS. Based on the evaluation performed on different hyperparameters , we found that keeping \textbf{p \& q both as 1}, and \textbf{number of walks and length of walks as 10} gives the best result.  Fig 5-9 shows the PR curve at different configuration of node2vec by varying a particular parameter, and Table II has AUC score for all the different configurations.

The parameters $p$ and $q$  play a crucial role in influencing the exploration dynamics of walks generated by the node2vec algorithm. Intuitively, they determine the speed at which walks explore and depart from the initial node, providing a flexible mechanism to interpolate between Breadth-First Search (BFS) and Depth-First Search (DFS) strategies.

Through a comprehensive evaluation involving various hyperparameter settings, we have identified that setting both\textbf{ $p$ and $q$  to 1, along with configuring the number of walks and the length of walks to 10}, yields optimal results in our study. This specific combination enhances the performance of the walk generation process.

The evaluation results are illustrated in Figures 5-9, showcasing Precision-Recall curves under different node2vec configurations by varying a specific parameter. Additionally, Table II provides the corresponding Area Under the Curve (AUC) scores for diverse configurations, affirming the superiority of the chosen parameter values (both) p and q as 1, with 10 walks and walk length). This empirical evidence underscores the effectiveness of the identified hyperparameter settings in achieving the best overall performance in the node2vec algorithm.

\begin{figure}[thpb]
      \centering
      \framebox{\parbox{3.2in}{\includegraphics[height=2.4in, width=3.2in]{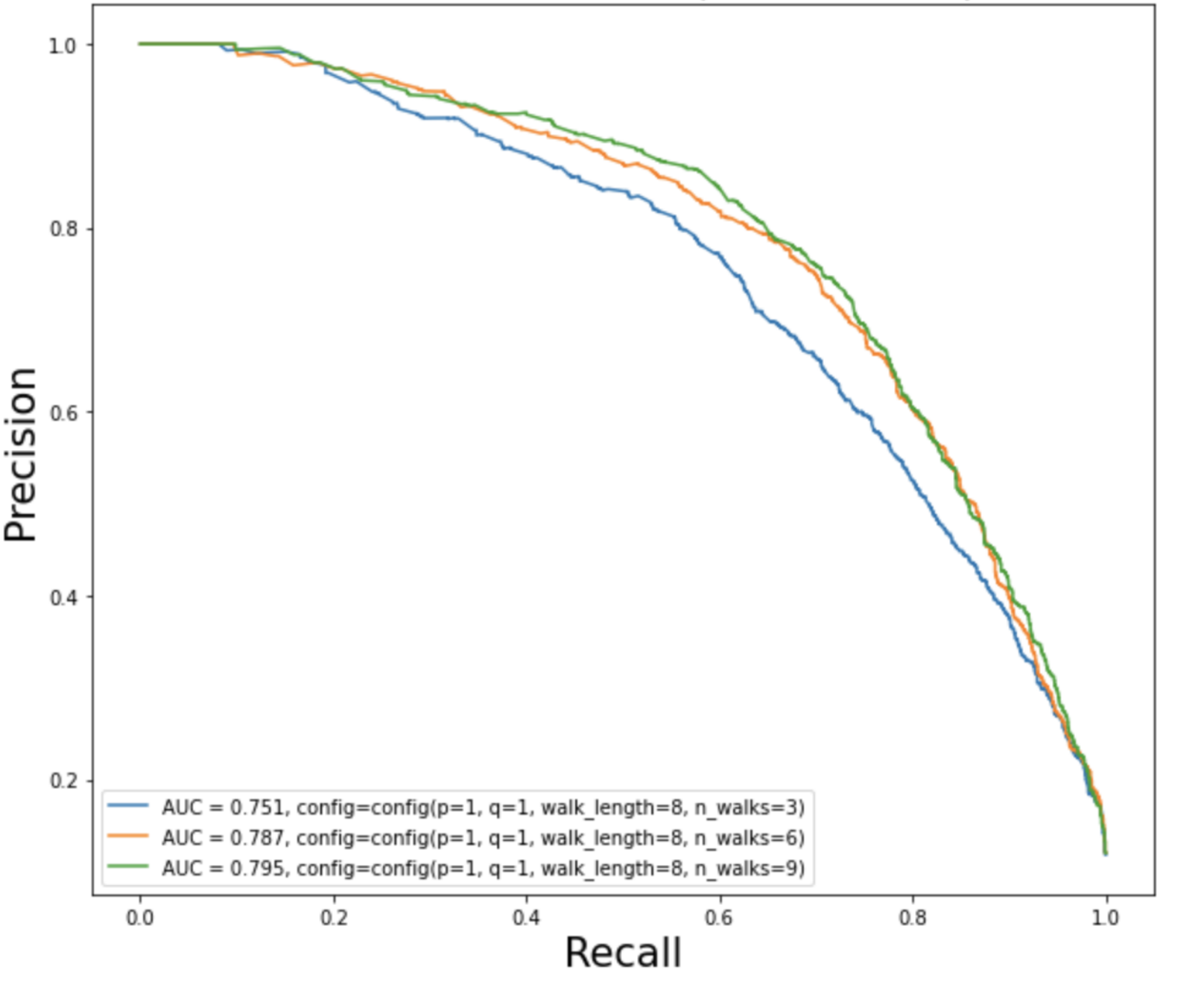}}}
      \caption{PR-Curve varying number of walks}
      \label{fig:5}
\end{figure}

\begin{figure}[thpb]
      \centering
      \framebox{\parbox{3.2in}{\includegraphics[height=2.4in, width=3.2in]{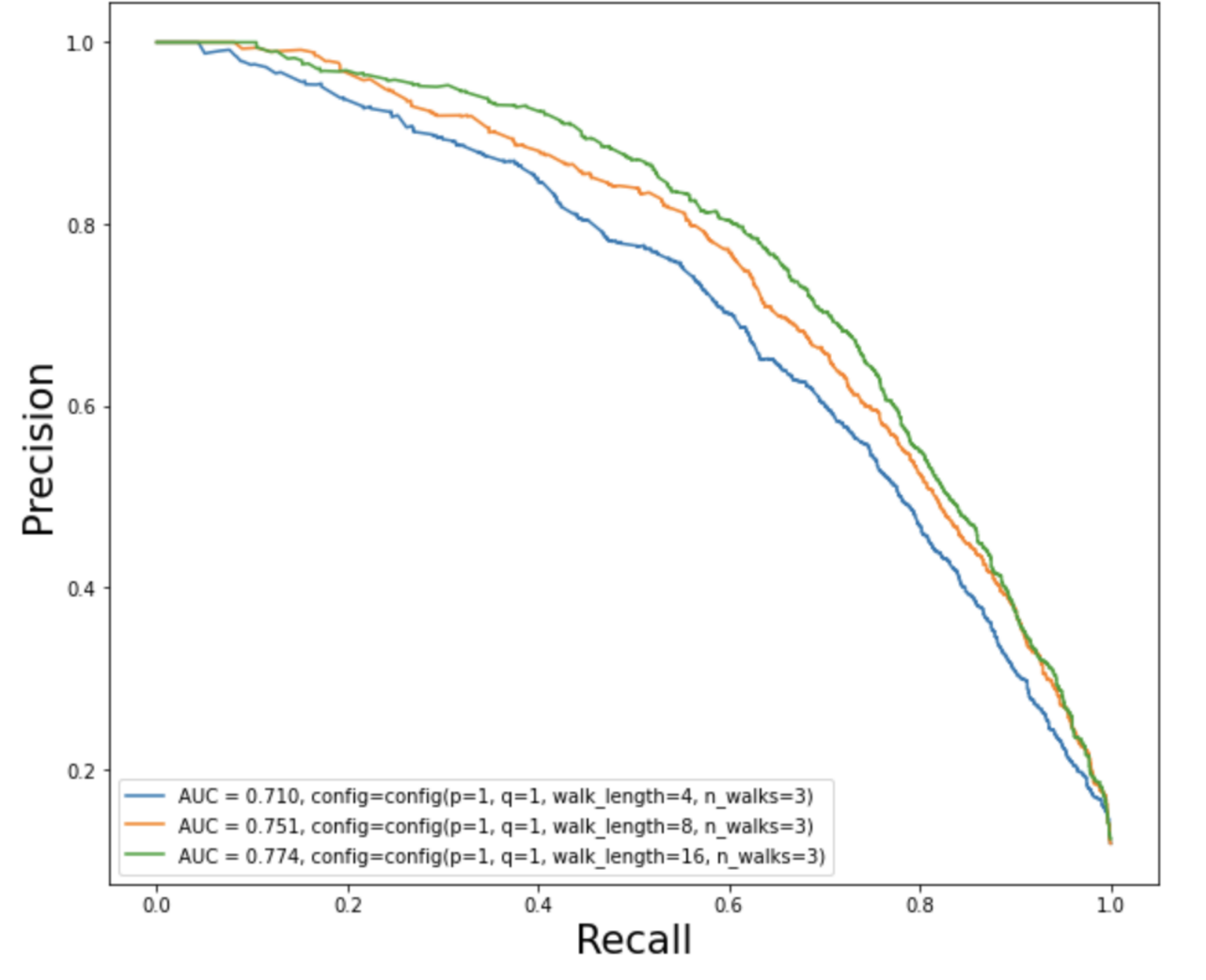}}}
      \caption{PR-Curve varying length of walks}
      \label{fig:6}
\end{figure}

\begin{figure}[thpb]
      \centering
      \framebox{\parbox{3.2in}{\includegraphics[height=2.4in, width=3.2in]{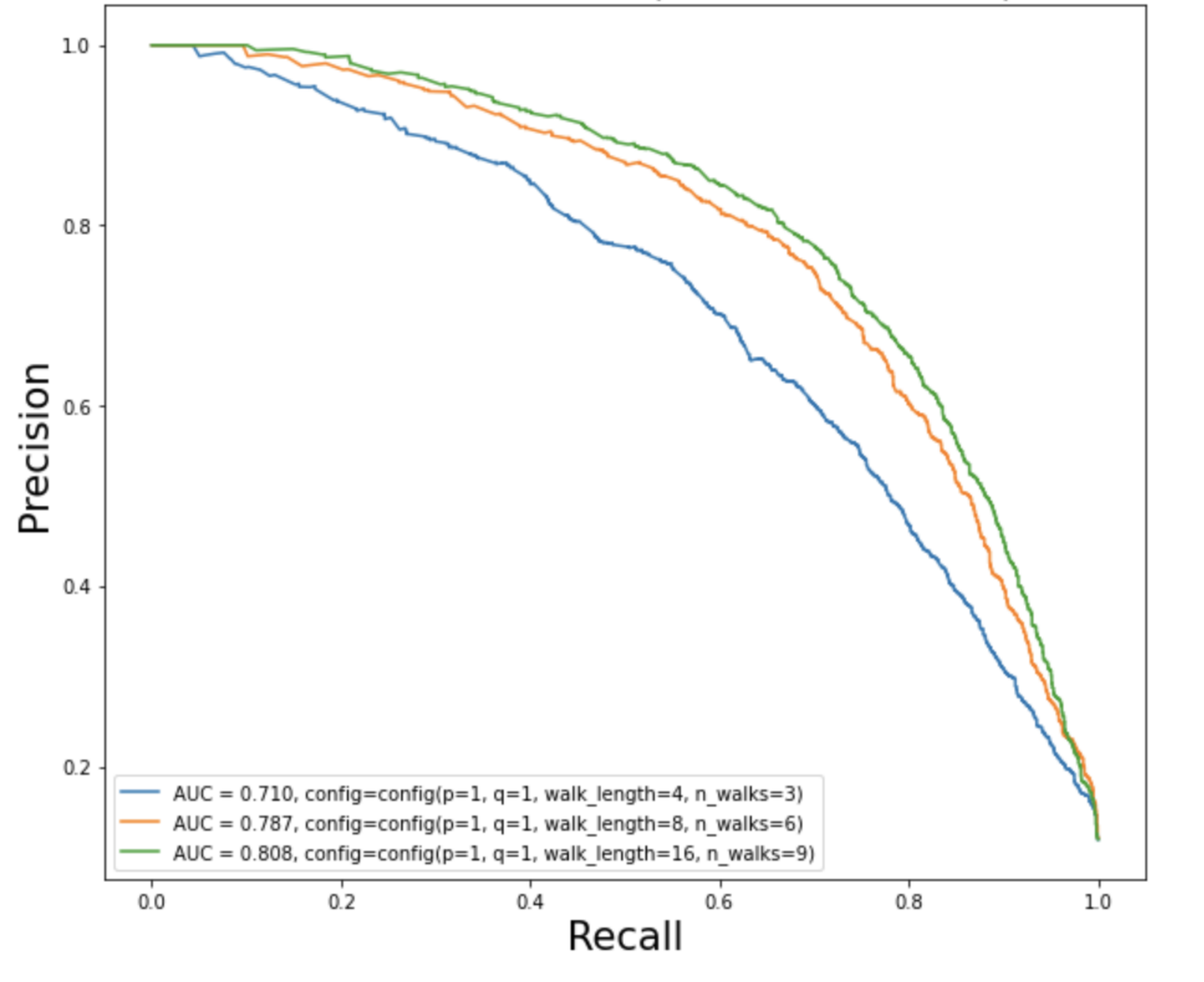}}}
      \caption{PR-Curve varying both number of walks and length of walks}
      \label{fig:7}
\end{figure}

\begin{figure}[thpb]
      \centering
      \framebox{\parbox{3.2in}{\includegraphics[height=2.7in, width=3.3in]{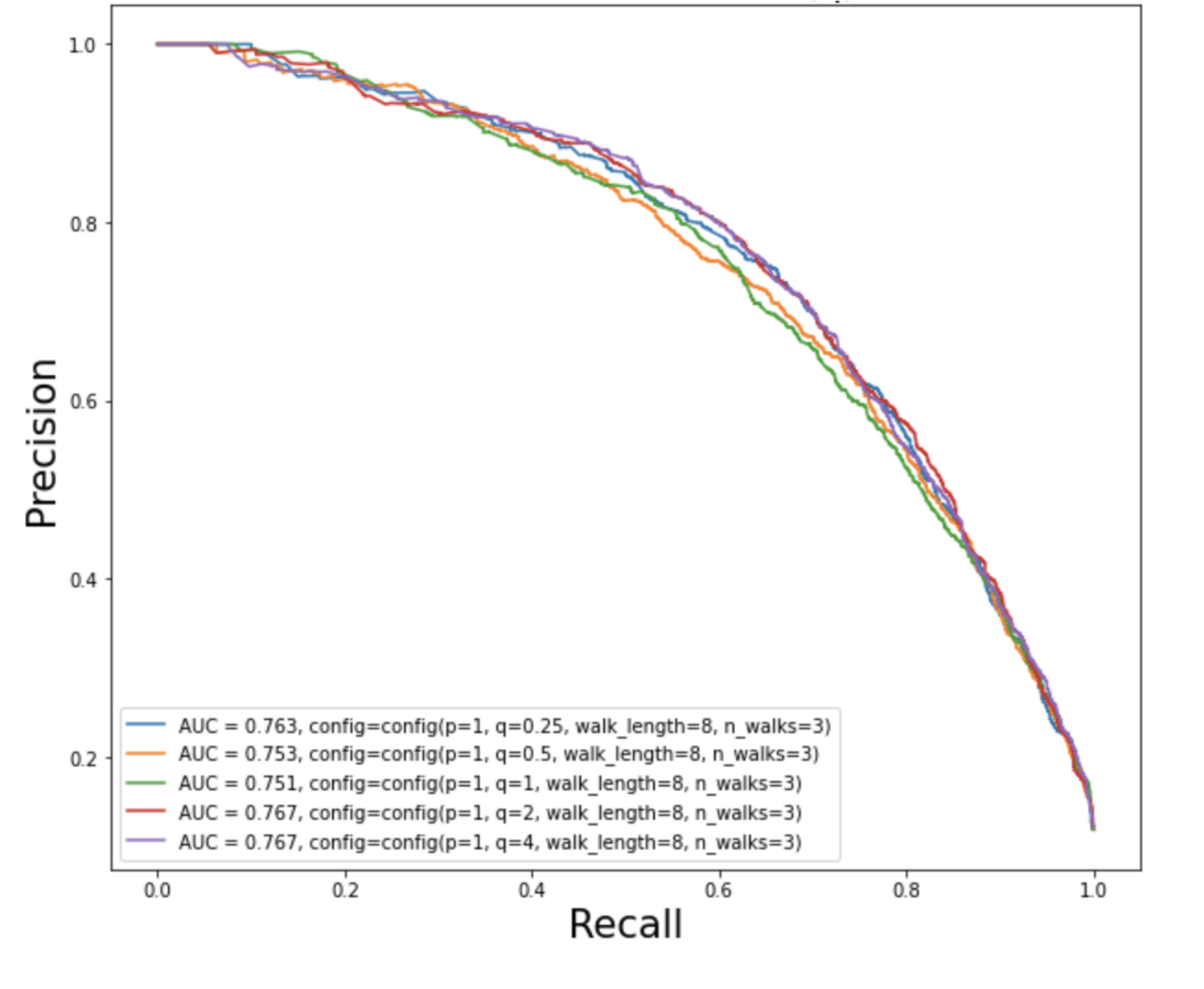}}}
      \caption{PR-Curve varying In-out Parameter (p)}
      \label{fig:8}
\end{figure}

\begin{figure}[thpb]
      \centering
      \framebox{\parbox{3.2in}{\includegraphics[height=2.7in, width=3.3in]{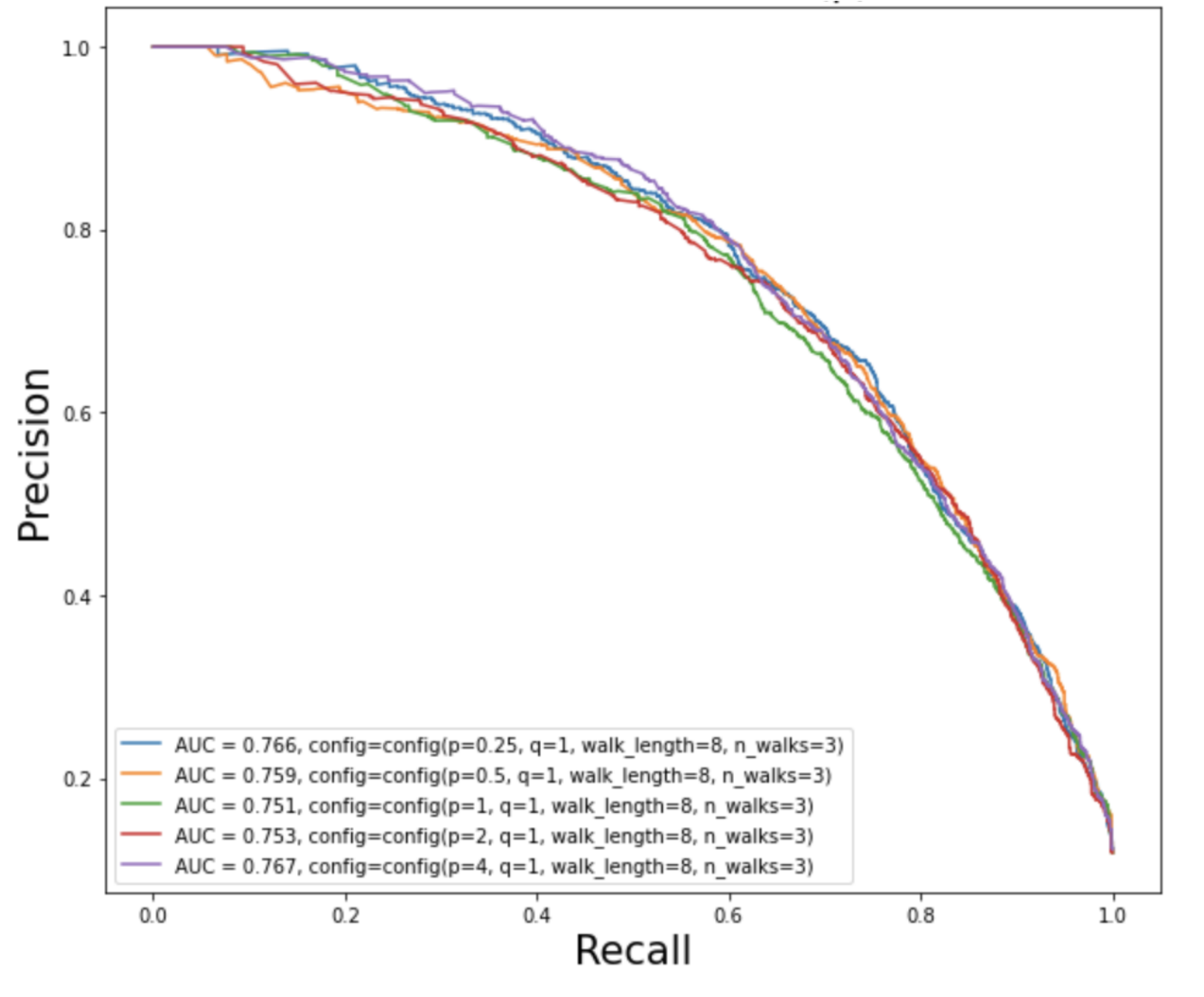}}}
      \caption{PR-Curve varying Return parameter (q)}
      \label{fig:9}
\end{figure}

 

\begin{table*}[t]
    \centering
    \renewcommand{\arraystretch}{1.4}
    \begin{tabular}{|c|c|c|c|c|} \hline 
         \textbf{Number of Walks} & \textbf{Length of Walk} & \textbf{Return Hyperparameter ($p$)}& \textbf{Inout Parameter ($q$)}& \textbf{AUC (Area Under Curve)}\\ \hline 
         \multicolumn{5}{|c|}{\textbf{Varying Number of Walks}}\\ \hline 
         3& 8& 1& 1&0.751\\ \hline 
         6& 8& 1& 1&0.787\\ \hline 
         9& 8& 1& 1&0.795\\ \hline 
         \multicolumn{5}{|c|}{\textbf{Varying Length of Walks}}\\ \hline 
 3& 4& 1& 1&\\ \hline 
 3& 8& 1& 1&0.751\\ \hline 
 3& 16& 1& 1&0.774\\ \hline 
 \multicolumn{5}{|c|}{\textbf{Varying both Number of Walks \& Length of Walks}}\\ \hline 
 3& 4& 1& 1&\\ \hline 
 6& 8& 1& 1&0.787\\ \hline 
 \textbf{10}& \textbf{10}& \textbf{1}& \textbf{1}&\textbf{0.814}\\ \hline 
 9& 16& 1& 1&0.808\\ \hline 
 \multicolumn{5}{|c|}{\textbf{Varying Return parameter ($p$)}}\\ \hline 
 3& 8& 0.25& 1&\\ \hline 
         3& 8& 0.5& 1&0.759\\ \hline 
         3& 8& 1& 1&0.751\\ \hline 
         3& 8& 2& 1&0.753\\ \hline 
         3& 8& 4& 1&0.767\\ \hline 
 \multicolumn{5}{|c|}{\textbf{Varying in-out parameter ($q$)}}\\ \hline 
 3& 8& 1& 0.25&\\ \hline 
 3& 8& 1& 0.5&0.753\\ \hline 
 3& 8& 1& 1&0.751\\ \hline
 3& 8& 1& 2&0.767\\\hline
 3& 8& 1& 4&0.767\\\hline
    \end{tabular}
    \renewcommand{\arraystretch}{1}
    \caption{Comparison based on varying hyper-parameters of \textit{DynN2VChain} Model}
    \label{table:2}
\end{table*}

We further trained two models completely till the latest blocks (16M blocks) with number of walks and length of walks as (3, 8) and (10, 10) respectively and observed that the node2vec model with hyperparam (10, 10) performs better than (3, 8).

\subsection{Evaluation on Ethereum Transaction Graph}

\begin{figure*}[ht]
    \centering

    \subfloat[PR-Curve with configuration (\textit{3 num\_walks, 8 walk\_length, 10M  initial\_eth\_blocks})]{%
        \includegraphics[width=.32\textwidth]{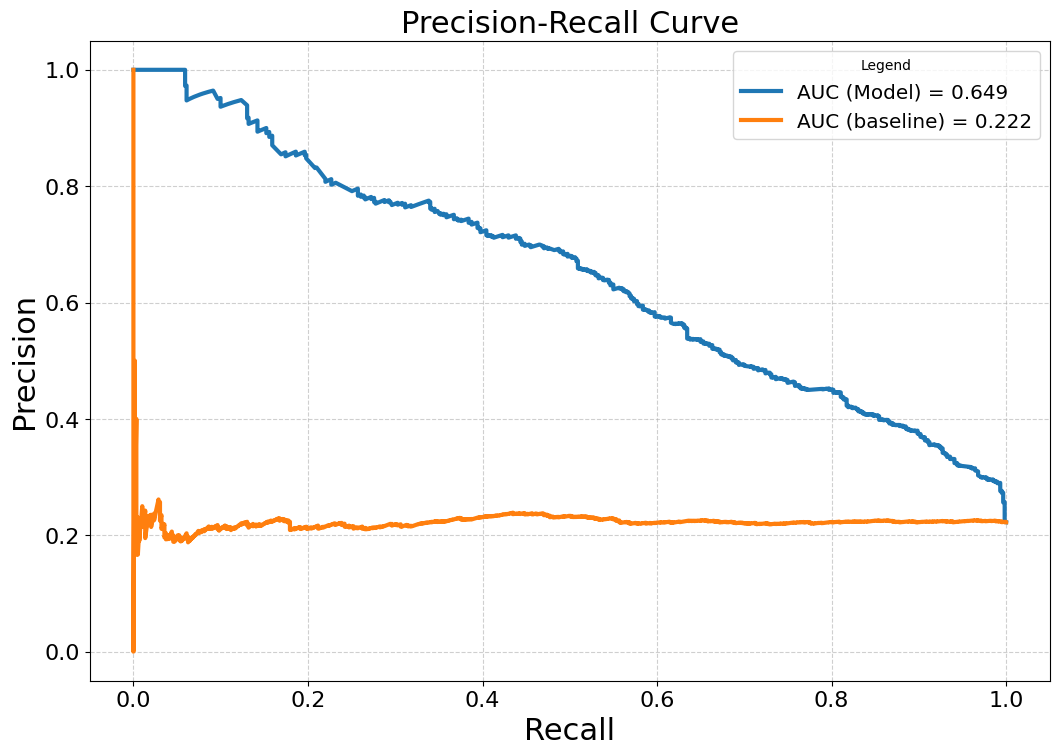} 
        \label{fig:subimage1}
    }
    \hfill
    \subfloat[PR-Curve with configuration (\textit{3 num\_walks, 8 walk\_length, 16M initial\_eth\_blocks})]{%
        \includegraphics[width=.32\textwidth]{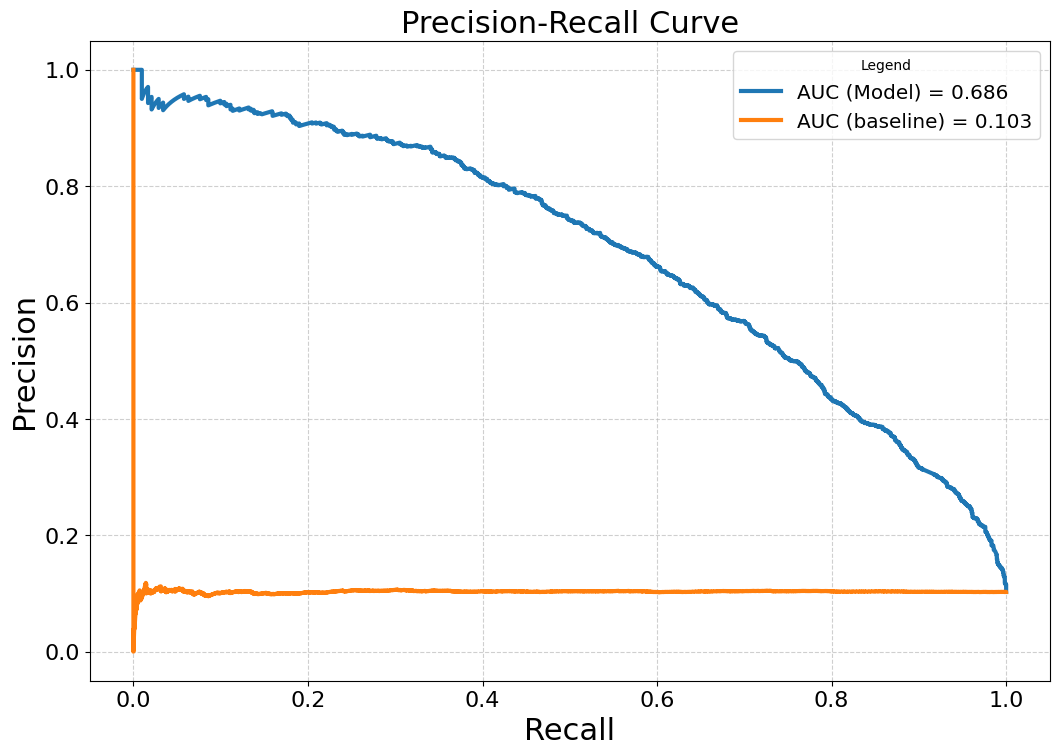} 
        \label{fig:subimage1}
    }
    \hfill
    \subfloat[PR-Curve with configuration (\textit{10 num\_walks, 10 walk\_length, 10M initial\_eth\_blocks})]{%
        \includegraphics[width=.32\textwidth]{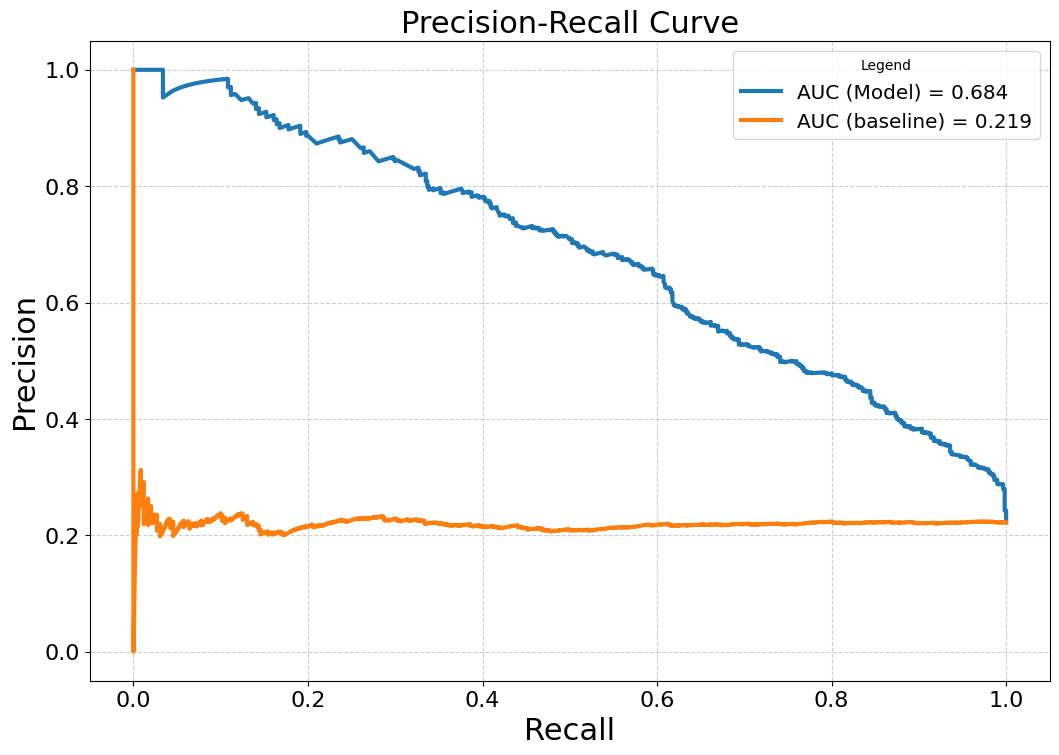} 
        \label{fig:subimage1}
    }
    \\[\smallskipamount]
    \centering
     \subfloat[PR-Curve with configuration (\textit{3 num\_walks, 8 walk\_length, 10M initial\_eth\_blocks})]{%
        \includegraphics[width=.32\textwidth]{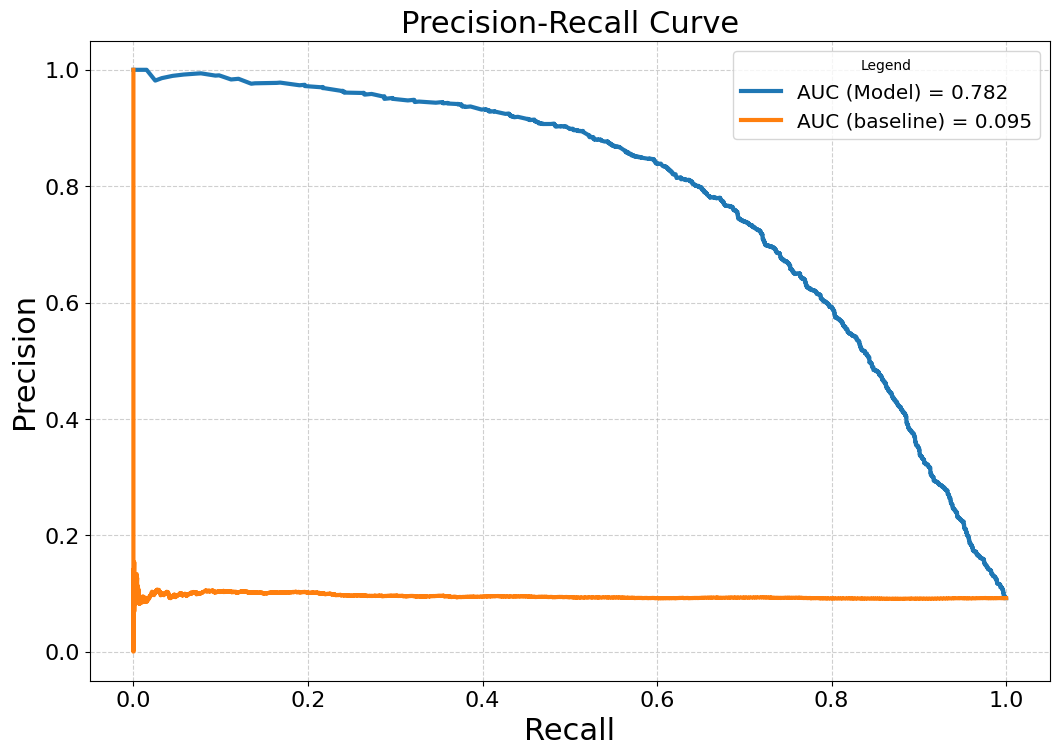} 
        \label{fig:subimage1}
    }
    \hfill
    \subfloat[PR-Curve with configuration (\textit{3 num\_walks, 8 walk\_length, 16M initial\_eth\_blocks})]{%
        \includegraphics[width=.32\textwidth]{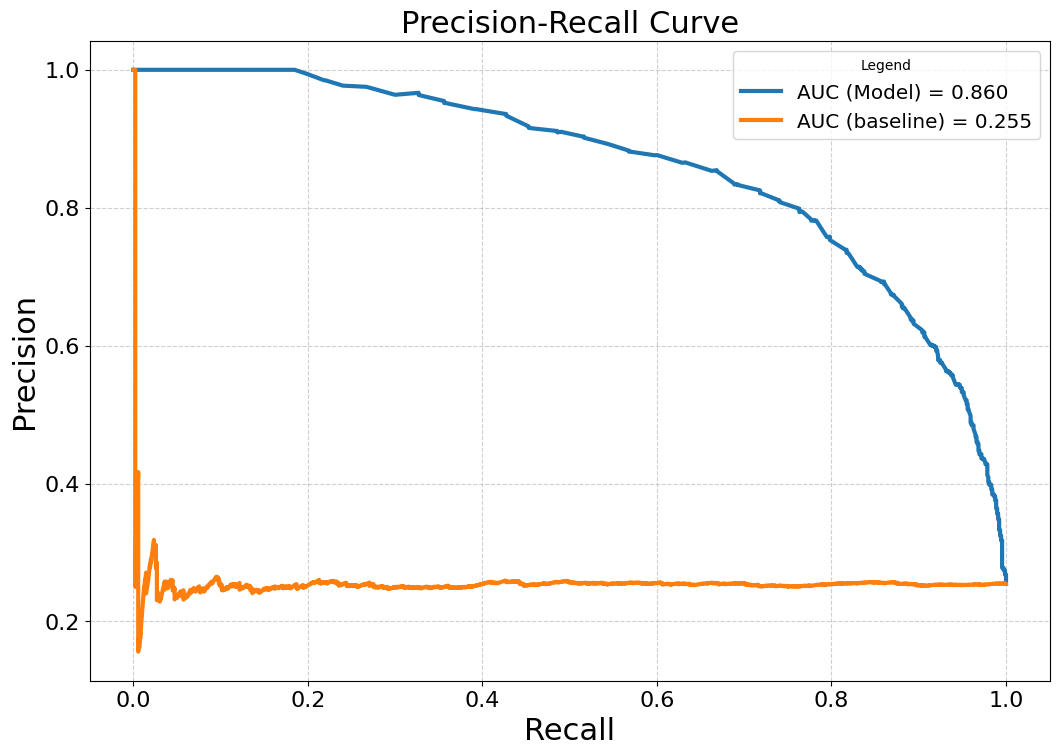}
        \label{fig:subimage1}
    }
    \hfill
    \subfloat[PR-Curve with configuration (\textit{10 num\_walks, 10 walk\_length, 10M initial\_eth\_blocks})]{%
        \includegraphics[width=.32\textwidth]{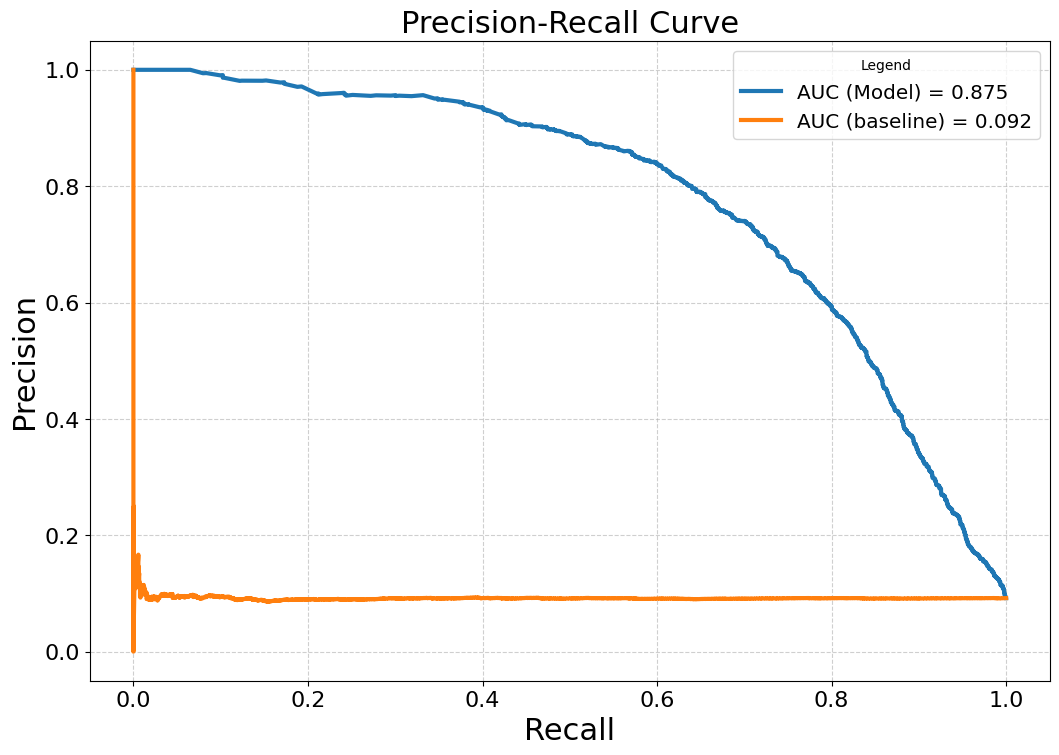}
        \label{fig:subimage1}
    }
    \\[\smallskipamount]


    \caption{Comparison of Embedding Generated from \textit{Propagation Methodology} and \textit{DynN2VChain} at different checkpoints}\label{fig:9}
\end{figure*}

\begin{table*}
\centering
\renewcommand{\arraystretch}{1.4}
\begin{tabular}{|c|c|c|c|}
\hline
\textbf{Embedding Type  /  Node2Vec Parameters ($num\_walks$ , $walk\_length$,  $initial\_eth\_blocks$)} & \textbf{(3, 8, 10)} & \textbf{(3, 8, 16)} & \textbf{(10, 10, 10)} \\
\hline
Embedding Propagation Methodology& 0.649 & 0.686 & 0.684 \\
\hline
DynN2VChain Embeddings & 0.782 & 0.860 & \textbf{0.875}\\
\hline

\end{tabular}
\caption{AUC Score for \textit{Embedding propogation} and \textit{DynN2VChain}}
\label{tab:3}
\renewcommand{\arraystretch}{1}
\end{table*}

We evaluated the embeddings generated by training a risk scoring model, and evaluating the model performance on a particular test-set by varying the hyperparameters of the dynamic node2vec model. 

For this evaluation, we have trained a random forest classifier with a fixed configuration and only the node2vec embeddings as an input feature to the model. We evaluated the embeddings by training the risk scoring and observing the overall performance of the risk scoring model on the test-set. Evaluation after training on all the blocks present in Ethereum chain would take very long for dynamic node2vec embeddings, instead we compared the model performance at different checkpoints like initial 5M, 8M, 10M and 16M blocks. 

We have compared the embeddings generated by dynamic node2vec embeddings at multiple checkpoints with the node2vec propagated embeddings generated on similar checkpoints with similar hyper-parameters. We observed \textbf{dynamic node2vec embeddings performs better than the node2vec propagated embeddings} in the task of generating risk scores. Fig. 10 and Table IV present comprehensive results obtained across different configuration of hyper-parameters, specifically across number of walks, walk length, and the initial Ethereum blocks considered in the transaction graph  for comparison of embedding propagation methodology and \textit{DynN2VChain} embeddings.

\section{Discussion and Conclusions}

In this paper, we introduced \textit{RiskSEA}, a scalable system for risk scoring of blockchain addresses, utilizing a rich feature set combining both behavioral and graph based features and employing a supervised machine learning model for efficient generation of normalized risk scores. We described the scalability challenges in generating graph based features posed by the vast number of Ethereum blockchain addresses. To address the scalability challenges,we proposed two approaches for generating node2vec embedding, an embedding propagation method and an incremental dynamic node2vec approach. The latter accommodates the dynamic nature of blockchain transaction graphs and introduces a novel horizontally scalable random walk technique for efficient incremental training. We performed a thorough analysis of the dynamic node2vec embeddings across various hyperparameters and also evaluated the approach which validates the computational scalability and effectiveness of the proposed approach. Our system handles the dynamic Ethereum network with approximately 275M nodes by generating a graph embedding for all the nodes present in the ethereum blockchain. This marks the first practical deployment of such a comprehensive system, addressing the evolving nature of blockchain transaction graphs.

Although we source the blockchain ground truth data from various sources, expanding the ground truth data should improve the risk scoring system. The efficacy of our approach is exemplified by the visualization of node2vec embeddings, illustrating distinct clusters of various class labels, as detailed in Appendix. This suggests the system's potential applications beyond risk scoring, that it can be utilized to predict other multi-class labels for blockchain addresses. It’s worth noting that our current focus on explainability for the generated risk scores using node2vec embeddings has been limited. Establishing a robust mechanism for explainability is imperative to offer a more comprehensive understanding of the results obtained from the system. Considering the expansion of our system, there is an exciting opportunity to broaden its scope to encompass other blockchains. This extension could potentially provide valuable insights into risk factors across different blockchain networks. Furthermore, it would be of interest to explore the feasibility of integrating our system with Layer2 chains, potentially unlocking a new dimension in risk assessment with the blockchain ecosystem.

\section{Acknowledgements}

Dr. Krishnamachari is a paid consultant for Coinbase and has assisted on this paper in this capacity. We would also like to thank Malar, Akshit and Rajat from ML team at Coinbase for their contributions in building components of the system, which were instrumental in conducting ablation studies.

\section{References}

\bibliographystyle{ieeetr}
\bibliography{Bibliography/citation}

\section{Appendix}

\subsection{Dynamic Node2Vec Embedding Evaluation on Ethereum Blockchain}

We have extracted data from various sources and addresses have categories and labels assigned to them. We randomly selected a set of addresses for each category and labels within each category. By plotting the embeddings in this reduced space, we were able to identify patterns and relationships among the addresses belonging to the same categories.

Fig 11-13 presents the T-SNE plots using \textit{DynN2VChain} embeddings on different address labels. The clear formation of clusters in figures of addresses belonging to a specific category instills confidence in the efficacy of the \textit{DynN2VChain} embedding generation approach. 

By capturing the inherent relationships and similarities among addresses, this feature enables us to discern meaningful patterns and groupings within the dataset. Leveraging the power of node2vec in our classification process holds great promise for accurately assigning addresses to their respective categories based on their embedded representations.

\begin{figure}[thpb]
      \centering
      \framebox{\parbox{3.5in}{\includegraphics[height=2.4in, width=3.5in]{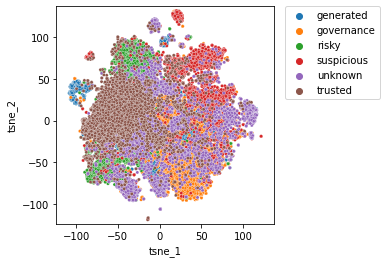}}}
      \caption{T-SNE Plot using DynN2VChain Embedding on all address labels}
      \label{fig:5}
\end{figure}

\begin{figure}[thpb]
      \centering
      \framebox{\parbox{3.5in}{\includegraphics[height=2.4in, width=3.5in]{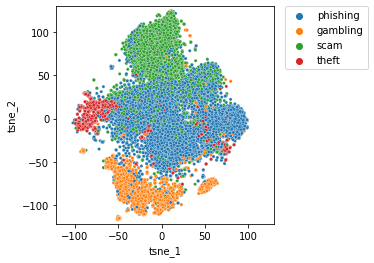}}}
      \caption{T-SNE Plot using DynN2VChain Embedding on address with risky labels}
      \label{fig:6}
\end{figure}

\begin{figure}[thpb]
      \centering
      \framebox{\parbox{3.5in}{\includegraphics[height=2.4in, width=3.5in]{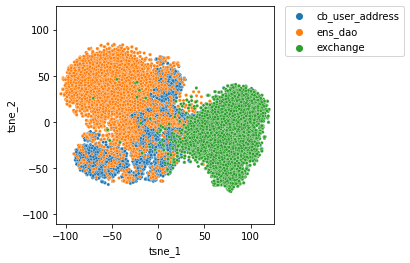}}}
      \caption{T-SNE Plot using DynN2VChain Embedding on address with trusted labels}
      \label{fig:7}
\end{figure}

\end{document}